\documentstyle[aps,preprint]{revtex}

\def\whatjournal{P}
\newlinechar=`\^^J

\def\ordernpb#1#2#3{{\bf#1} (#3) #2}
\if P\whatjournal {\global\def\order#1#2#3{\orderprd{#1}{#2}{#3}}}
                    \immediate\write16{^^J PRD references ^^J}\else 
                   {\global\def\order#1#2#3{\ordernpb{#1}{#2}{#3}}}
                    \immediate\write16{^^J NPB references ^^J} 
\fi

\def\app#1#2#3{{\it Acta Phys. Pol. {\bf B}}\order{#1}{#2}{#3}}
\def\arnps#1#2#3{{\it Ann. Rev. Nucl. Phys. Sci. }\order{#1}{#2}{#3}}
\def\ijmpa#1#2#3{{\it Int. J. of Mod. Phys. {\bf A}}\order{#1}{#2}{#3}}

\def\npb#1#2#3{{\it Nucl. Phys. {\bf B}}\order{#1}{#2}{#3}}

\def\plb#1#2#3{{\it Phys. Lett. {\bf B}}\order{#1}{#2}{#3}}

\def\ppnp#1#2#3{{\it Prog. Part. Nucl. Phys.\ }\order{#1}{#2}{#3}}
\def\pr#1#2#3{{\it Phys. Rev.\ }\order{#1}{#2}{#3}}

\def\prl#1#2#3{{\it Phys. Rev. Lett.\ }\order{#1}{#2}{#3}}

\def\prd#1#2#3{{\it Phys. Rev. {\bf D}}\order{#1}{#2}{#3}}

\def\acal{{\cal A}}
\def\lcal{{\cal L}}
\def\ocal{{\cal O}}
\def\sm{Standard Model}
\def\etal{{\it et.al.}}
\title{On effective interactions containing
$ \partial \cdot W^\pm $ and $ \partial \cdot Z $ factors.}

\author{Martin B Einhorn~\footnote{meinhorn@umich.edu}}
\address{The Randall Lab. of Physics \\ University of Michigan\\ Ann Arbor, MI
48109, U.S.A.}
\author{Jos\'{e} Wudka~\footnote{jose.wudka@ucr.edu}}
\address{Department of Physics \\ University of California, Riverside \\
Riverside, CA 92521-0413 U..S.A.}
\preprint{UCRHEP-T171 \  UM-HEP-96-12}
\date{\today}
\begin{document}

\maketitle

\begin{abstract}
The effects of effective interactions containing the factors $ \partial
\cdot W^\pm $ and $ \partial \cdot Z $ is studied within a consistent
effective Lagrangian formalism. It is shown that such terms are
redundant
\end{abstract}


\bigskip
\bigskip

One of the most important goals of current and future high-energy
experiments is to determine the structure of the physics underlying the
\sm. Given the possibility that non-standard effects may not be directly
observable, significant attention has been paid to the various virtual
effects generated by the heavy physics~\cite{phen}. When
seeking a model-independent description of such virtual effects,
the optimal approach is based on a gauge-invariant effective
Lagrangian~\cite{reviews}.

It might be puzzling to require the effective lagrangian to be gauge
invariant since, at least in the case of gauge theories, the explicit
derivation of the effective action requires that the underlying theory
be gauge-fixed in some way. This apparent contradiction is solved by
noting that one can use a background field gauge fixing method~\cite{bgf}
which generates a gauge invariant effective action in the light fields.

The usual parameterization of the effective interactions, however, regularly
ignores terms containing the divergence of the vector
fields~\cite{Hagiwara.et.al}.  This has no effect whenever the vector field is
coupled to a conserved current or when the fermion and scalar masses can be
ignored. But the recent determination of the top mass~\cite{top} raises the
possibility that effective interactions containing $ \partial \cdot W^\pm $ or
$ \partial \cdot Z $ might be of phenomenological interest, because the top has
non-negligible branching ratio into processes involving virtual $W$ bosons.  In
this paper we will consider the effects and observability of such terms. We
will see that terms containing such divergences are
redundant ({\it i.e.} equivalent to other effective interactions) and
do not produce {\it any} physical effects.

In proving this assertion we will use the fact that the S-matrix is
invariant under field redefinitions~\cite{redef}, and the
fact that operators which vanish upon use of the equations of motion do
not contribute to the S-matrix~\cite{eom}. We will see that any operator
containing the divergence of a vector field can be rewritten as the sum {of}
two terms, one which does not contain such divergences, plus another
which vanishes when the equations of motion are imposed.

It must be noted that the terms under consideration can be
eliminated completely in the Landau gauge. But in this case the
would-be Goldstone bosons are massless and
must be included in the effective Lagrangian,
and the problem is shifted to the terms containing these unphysical
excitations.\footnote{In particular, the decoupling theorem is problematic in
the Landau gauge.}

The effective interactions generated by the heavy physics correspond to
a series of operators in the light fields. Self-consistency requires
that they respect the same gauge symmetries as the \sm.
In presenting the proof we will consider a general theory containing
gauge-bosons, fermions and scalars, the scalar terms may or may not be
polynomial (thus including the case where the scalars appear in unitary
fields). We will not need to specify the group or particle content of the
low energy theory. Gauge fields will be denoted by $ A_\mu^a $, we
collect all fermions in a multiplet $ \psi $ and all scalars (if
present) in a {\it real} multiplet $ \chi $. This is always possible
though $ \psi $ and $ \chi $ will transform, in general, under reducible
representations of the gauge group.

For purposes of field redefinition, we must consider the ``classical''
equations of motion.  For the fermions and
scalars {these} are
\begin{eqnarray}
\label{eq:eom}
&\not \!\! D \psi = Q \psi &\\ \nonumber
& D_{ik}^\mu \left[ g_{ij}(\chi) \left( D_\mu \chi \right)_j \right] =
Z_i & \nonumber
\end{eqnarray}
where $Q$ contains all the mass terms and Yukawa couplings in the
theory, $Z_j$ contains the potential terms and fermions couplings, and
$g$ denotes the metric in the scalar sector, defined through the
kinetic term
\begin{equation}
\lcal_{\chi; {\rm kin}} = {1\over2} g_{ i j } ( \chi) \left( D^\mu \chi
\right)_i
\left( D^\mu \chi \right)_j .
\end{equation}
The covariant derivatives can always be written $ D_\mu = \partial_\mu +
A_\mu $, $ A_\mu = \sum_a g_a T^a A_\mu^a $, where $ \{ T^a \} $ denotes the
anti-Hermitian generators of the relevant representation, and $ g_a $ the
gauge coupling constants with $ g_a = g_b $ whenever $a$ and $b$ belong
to the same semi-simple or $U(1)$ factor in the gauge group.
Using (\ref{eq:eom}) we obtain
\begin{eqnarray}
\label{eq:eom1}
D^2 \psi &=& \left[ {1 \over 2 i }  \sigma^{ \mu \nu } F_{ \mu \nu } + Q
\right] \psi \\ \nonumber
D^2 \chi_i &=& g^{-1}_{ij} \left[ Z_j - (D^\mu g)_{jk} (D_\mu \chi)_k
\right]
\end{eqnarray}
where $g^{-1}$ denotes the inverse of $g$.

Consider an effective operator containing a factor $ \partial \cdot
A^a $, its general form is~\footnote{We assume that appropriate integration by
parts have been performed so that no derivatives act on $ \partial \cdot
A $.}
\begin{equation}
\ocal = X^a \left( \partial \cdot A^a \right) + Y .
\end{equation}
Due to gauge invariance, ordinary derivatives must be part of covariant
derivatives and $A$ arises only as part of a
covariant derivative (field strengths can be written as commutators
of covariant derivatives). It follows that $ \ocal $ can be
rewritten as $ Y' + \mbox{tr} \left\{ W_1 D^2 W_2 \right\} $ where the
trace is over group indices, $Y'$ is gauge invariant, and $W_{1,2} $
represent some combination of fields. Using now (\ref{eq:eom},\ref{eq:eom1}) we
find that we can always rewrite this as an operator which does not contain a
$ \partial \cdot A $ factor plus another operator which vanishes
whenever the classical equations of motion are valid (and which does contain
the divergence terms). Thus we can write
\begin{equation}
\ocal = \ocal' + \sum_\phi \acal_\phi { \delta S \over \delta \phi }
\end{equation}
where the summation is over all the fields appearing in $W_2$ (denoted
collectively by $ \phi $), $S$ denotes the classical action, and $\ocal'$
comprises all terms stemming from $Y'$ and the replacement of $ D^2 W_2$
using the equations of motion ({\it i.e.} all terms containing the
right-hand side of (\ref{eq:eom1})).

It is well known~\cite{eom}, however, that the S-matrix generated by a
Lagrangian containing $ \ocal $ is the same as the one generated by a
Lagrangian containing $ \ocal' $. Since the latter operator does not
contain a $ \partial \cdot A $ factor, the claim made above is proved.
Note also that operators containing several $ \partial \cdot A $ factors
can be treated iteratively, eliminating one factor at a time by the
above procedure.

%

\end{document}